\documentstyle[preprint,tighten,eqsecnum,aps]{revtex}
\begin{document}
\preprint{TRI-PP-94-65}
\draft
\title{Compton scattering by a pion and off--shell effects}
\author{S.\ Scherer\thanks{Address after Sept.\ 1, 1994:
Institut f\"ur Kernphysik, Johannes Gutenberg--Universit\"at,
J.\ J.\ Becher--Weg 45, D--55099 Mainz, Germany.} and H.\ W.\ Fearing}
\address{TRIUMF, 4004 Wesbrook Mall, Vancouver, B.\ C.,\\ Canada V6T 2A3}
\date{August 15, 1994}
\maketitle
\begin{abstract}
We consider Compton scattering by a pion in the framework of chiral
perturbation theory.
We investigate off--shell effects in the s-- and u--channel pole diagrams.
For that purpose we perform a field transformation which, in comparison with
the standard Gasser and Leutwyler Lagrangian, generates additional terms at
order $p^4$ proportional to the lowest--order equation of motion.
As a result of the equivalence theorem the two Lagrangians predict the
same Compton scattering S--matrix even though they generate different
off--shell form factors.
We conclude that off--shell effects are not only model--dependent but
also representation--dependent.
\end{abstract}
\narrowtext
\section{Introduction}
The question of how to incorporate effects due to the `internal' structure of
particles which are not on their mass shell has a long history.
As early as 1954, in their pioneering work on low--energy Compton scattering
by spin--1/2 particles,
Gell--Mann and Goldberger \cite{Gell--Mann} took into account the fact
that the electromagnetic vertex of the nucleon involving off--shell
momenta is more complicated than the operator for asymptotically
free states.
In fact, the most general expression for the $\gamma NN$
vertex operator contains 12 invariant functions (`form factors')
depending on three scalar variables \cite{Bincer}.
Clearly, similar observations hold for any vertex function.
For example, the electromagnetic vertex of an off--shell pion contains two
invariant functions which depend on the squared four--momentum transfer,
and the invariant masses associated with the initial and final
legs of the vertex \cite{Nishijima}.
This has to be compared with {\em one} phenomenological form factor for
{\em free} particles depending on {\em one} scalar variable, namely,
the squared momentum transfer.

A considerable amount of work has been devoted to attempts to obtain
quantitative predictions for the off--shell behavior of both the
$\gamma NN$ and the $\gamma\pi\pi$ vertex using either dispersion
theory \cite{Bincer,Nyman1,Bos1} or calculations in microscopic
models \cite{Naus,Tiemeijer,Song,Bos2,Rudy,Weiss,Doenges}.
The importance of a consistent treatment of off--shell functions,
in particular in the context of gauge invariance, has been realized
and emphasized by many authors \cite{Naus,Gross,Singh,Ohta,Bos3,Frank}.
In this context the most important constraint is provided by the
Ward--Takahashi identity \cite{Ward,Takahashi}.
Even though the need for microscopic models to determine off--shell
effects has been repeatedly stressed, very little has been said
about whether it is, even in principle, possible to obtain information about
off--shell contributions from experimental data.

A relatively simple process to address and clarify the issues of off--shell
effects is Compton scattering by a spin--0 particle, e.g., by the pion.
Following the method of Gell--Mann and Goldberger \cite{Gell--Mann}
the invariant amplitude may be divided into two classes, A and B, where class
A essentially contains the pole terms and class B the rest.
A calculation of the s-- and u--channel pole terms involves an intermediate
off--shell pion and thus contributions from the off--shell electromagnetic
vertex to the scattering amplitude.
One would expect that a microscopic calculation in the framework of a
well--defined model should thus allow for a unique identification of
such contributions.
Furthermore, provided the calculated scattering amplitude was in good agreement
with experimental data, one would indirectly obtain the off--shell form factor
from the model.
However, we will show that this is, in fact, {\em not} the case.

For this purpose we will study the Compton scattering amplitude in the
framework of chiral perturbation theory \cite{Gasser1,Gasser2}.
We will discuss two {\em equivalent} forms of the chiral Lagrangian which can
be related by an appropriate field redefinition.
In other words, we will deal with the {\em same} microscopic model
in different representations.
Using these two Lagrangians, we then calculate the Compton scattering
amplitude up to $O(p^4)$ in the momentum expansion.
Even though both Lagrangians yield the same result for the Compton scattering
amplitude they provide different predictions for the off--shell form factors.
This clearly shows that off--shell form factors are both, model-- and
representation--dependent, and thus even the 'correct' microscopic model
would not give a unique result for off--shell form factors.

\section{Change of field variables in the chiral Lagrangian}
The effective Lagrangian of chiral perturbation theory \cite{Gasser1,Gasser2}
provides the most general description of the interaction between the
low--energy degrees of freedom of the strong interaction, namely the Goldstone
bosons ($\pi,K,\eta$) of spontaneous $SU(3)_L\times SU(3)_R$ chiral symmetry
breaking, as well as their interaction with external fields.
It is organized as an expansion in powers of covariant derivatives of the
fields, scalar and pseudoscalar external sources (containing the quark mass
terms), and field strength tensors,
\begin{equation}
\label{clag}
{\cal L}={\cal L}_2+{\cal L}_4 + \dots,
\end{equation}
where the subscripts refer to the order in the momentum expansion.
Covariant derivatives which, e.g., may contain the electromagnetic field
are counted as $O(p)$, external sources and field strength tensors
are $O(p^2)$.
Beyond lowest order in the momentum expansion, use of the `classical equation
of motion', i.e., the one derived from the lowest--order Lagrangian
${\cal L}_2$, is made to minimize the number of independent structures
at a given order.
For example, at $O(p^4)$ two terms have been eliminated \cite{Gasser1,Gasser2}.
This procedure can be interpreted in terms of field transformations
\cite{Georgi,Leutwyler,Scherer}.
The use of field transformations as a means to simplify interaction terms
is motivated by the equivalence theorem \cite{Haag,Kamefuchi,Coleman}.
Lagrangians which are related by field transformations generate
the same on--shell S--matrix elements, provided the transformation
satisfies certain properties.
The extension to effective Lagrangians has been discussed by several authors
\cite{Georgi,Arzt,Grosse--Knetter}.

In the following, we will use the freedom of field transformations to define
two equivalent forms of the chiral Lagrangian at $O(p^4)$.
We will apply the two Lagrangians to a computation of the Compton scattering
amplitude at $O(p^4)$ and investigate the role of off--shell form factors.
For this purpose we have to provide a short discussion of the concept of field
transformations in the framework of the chiral Lagrangian
(for more details, see Refs.\
\cite{Georgi,Leutwyler,Scherer,Coleman,Arzt,Grosse--Knetter}).

The most general chiral Lagrangian at $O(p^2)$ is given by \cite{Gasser2}
\begin{equation}
\label{l2}
{\cal L}_2 = \frac{F_0^2}{4} Tr \left ( D_{\mu} U (D^{\mu}U)^{\dagger} \right)
+\frac{F_0^2}{4} Tr \left ( \chi U^{\dagger}+ U \chi^{\dagger} \right ),
\quad U(x)=\exp\left( i\frac{\phi(x)}{F_0} \right ),
\end{equation}
where $\phi(x)$ is a traceless, hermitian $3\times 3$ matrix containing
the eight Goldstone bosons:
\begin{equation}
\label{phi}
\phi = \left (
\begin{array}{ccc}
\pi^0+\frac{1}{\sqrt{3}}\eta & \sqrt{2} \pi^+ & \sqrt{2} K^+ \\
\sqrt{2} \pi^- & -\pi^0+\frac{1}{\sqrt{3}}\eta & \sqrt{2} K^0 \\
\sqrt{2} K^- & \sqrt{2} \bar{K}^0 & -\frac{2}{\sqrt{3}}\eta
\end{array}
\right ).
\end{equation}
The covariant derivative $D_\mu U$ is defined as
\begin{equation}
\label{covder}
D_\mu U=\partial_\mu U -i R_\mu U +iU L_\mu,
\end{equation}
where $R_\mu$ and $L_\mu$ are hermitian, traceless $3\times 3$ matrices
containing
the 16 gauge fields associated with local $SU(3)_L\times SU(3)_R$ invariance.
$F_0$ is the pion decay constant in the chiral limit ($F_0\approx 93 MeV$),
and $\chi$, eventually, contains the quark masses (for further details see
Refs.\ \cite{Gasser1,Gasser2}).

The field variables $\phi$ of Eq.\ (\ref{phi}) may be expressed as a function
of another set of field variables $\phi'$ (and their derivatives).
The transformation is most conveniently represented by relating the $SU(3)$
matrix $U(x)$ of Eq.\ (\ref{l2}) to another $SU(3)$ matrix $U'(x)$,
\begin{equation}
\label{phichi}
U(x)=\exp\left( i\frac{\phi(x)}{F_0} \right )
\equiv\exp(iS(U'))U'(x)=\exp(iS(U'))\exp\left( i\frac{\phi'(x)}{F_0}
\right ),
\end{equation}
where $S=S^{\dagger}$, and $Tr(S)=0$.
The generator $S(U')$ has to satisfy certain requirements \cite{Scherer}
which guarantee that $U'$ has the proper transformation behavior under
the group $SU(3)_L\times SU(3)_R$, parity, and charge conjugation.
At $O(p^2)$ two such generators exist \cite{Scherer}:
\begin{equation}
\label{generators}
S_2(U')=i\alpha_1(D^2U'U'^{\dagger}-U' (D^2U')^{\dagger})
+i\alpha_2\left(\chi U'^{\dagger}-U' \chi^{\dagger}-\frac{1}{3}
Tr(\chi U'^{\dagger}-U'\chi^{\dagger})\right),
\end{equation}
where $\alpha_1$ and $\alpha_2$ are arbitrary real parameters.
Inserting Eq.\ (\ref{phichi}) into Eq.\ (\ref{l2}), the result can be
written as
\begin{equation}
\label{l2n}
{\cal L}_2(U) = {\cal L}_2(U')+\delta^{(1)} {\cal L}_2(S,U')
+\delta^{(2)} {\cal L}_2(S,U')+\dots,
\end{equation}
where the superscript denotes the power of $S$ in the corresponding expression.
For the present purpose, namely for constructing an equivalent Lagrangian up to
$O(p^4)$, we can neglect all terms $\delta^{(n)}{\cal L}_2$
in Eq.\ (\ref{l2n})
with $n \ge 2$, since they are at least $O(p^6)$, and replace $S\rightarrow
S_2$ in $\delta^{(1)}{\cal L}_2$.
Up to a total divergence one finds for $\delta^{(1)}{\cal L}_2$
\begin{equation}
\label{dl2}
\delta^{(1)} {\cal L}_2(S,U')= \frac{F^2_0}{4} Tr(iS(U')
{\cal O}_{EOM}^{(2)}(U')),
\end{equation}
where
\begin{equation}
\label{eom2}
{\cal O}^{(2)}_{EOM}(U')=D^2 U' U'^\dagger - U' (D^2 U')^\dagger
-\chi U'^\dagger +U' \chi^\dagger
+\frac{1}{3}Tr\left(\chi U'^\dagger-U'\chi^\dagger\right)
\end{equation}
has the functional form of the classical equation of motion derived from
${\cal L}_2(U')$.
However, we do {\em not} assume Eq.\ (\ref{eom2}) to vanish.
Note that $\delta^{(1)} {\cal L}_2(S_2,U')$ is of $O(p^4)$, since both the
classical equation of motion and $S_2$ are of $O(p^2)$.
Furthermore, when inserted into the $O(p^4)$ Lagrangian of
Gasser and Leutwyler,
${\cal L}_4^{GL}$
(see Eq.\ (6.16) of Ref.\ \cite{Gasser2}), the field redefinition
induces a change at $O(p^6)$ which we are not interested in here.
Thus by a simple transformation of the interpolating field we generate an
infinite set of {\em equivalent } Lagrangians depending on two parameters
$\alpha_1$ and $\alpha_2$.

Setting $\alpha_1=4\beta_1/F^2_0$ and $\alpha_2=-4(\beta_1+\beta_2)/F^2_0$
and renaming $U'\rightarrow U$, one can bring Eq.\ (\ref{dl2}) into the
completely equivalent form chosen in Ref.\ \cite{Rudy},
\begin{equation}
\label{loff}
\delta^{(1)}{\cal L}_2(S_2,U) \equiv
\Delta{\cal L}_4(U)=\beta_1 Tr({\cal O}_{EOM}^{(2)}(U)
{\cal O}^{(2)\dagger}_{EOM}(U))
+\beta_2Tr((\chi U^\dagger-U\chi^\dagger){\cal O}_{EOM}^{(2)}(U)).
\end{equation}
In Ref.\ \cite{Rudy} it was explicitly shown that Eq.\ (\ref{loff})
results in contributions to the electromagnetic vertex which appear
only when one of the legs is off shell.

\section{Application to Compton scattering}
In this section we want to compare the predictions for the Compton scattering
amplitude up to $O(p^4)$ in the framework of the two Lagrangians
\begin{eqnarray}
\label{ls}
{\cal L}&=&{\cal L}_2+{\cal L}_4^{GL},\nonumber\\
{\cal L}'&=&{\cal L}_2+{\cal L}_4^{GL}+\Delta{\cal L}_4,
\end{eqnarray}
where ${\cal L}^{GL}_4$ and $\Delta {\cal L}_4$ denote the standard Lagrangian
of Gasser and Leutwyler \cite{Gasser2}, and  the additional term of
Eq.\ (\ref{loff}), respectively.
For that purpose the electromagnetic field and the quark masses are introduced
into the Lagrangians of Eq.\ (\ref{ls}) by making the replacement
\begin{equation}
\label{id}
R_\mu = L_\mu =-e Q A_\mu,\quad \chi=2B_0 M,
\end{equation}
where $eQ$ ($e>0$) and $M$ are the quark charge and mass matrices,
respectively,
\begin{equation}
\label{qandm}
Q=\left(\begin{array}{ccc} \frac{2}{3}&0&0\\0&-\frac{1}{3}&0\\
0&0&-\frac{1}{3}\end{array}\right),\quad
M=\left(\begin{array}{ccc} m_u&0&0\\0&m_d&0\\
0&0&m_s\end{array}\right),
\end{equation}
and $B_0$ is a phenomenological constant related to the scalar quark
condensate.
Furthermore, at lowest order in the momentum expansion one finds
$M_{\pi^+}^2=(m_u+m_d)B_0$ \cite{Gasser2}.
According to the power counting scheme of chiral perturbation theory
(for details see, e.g., Ref.\ \cite{Donoghue1}) the additional term
$\Delta{\cal L}_4$ will contribute at $O(p^4)$ to the Compton scattering
amplitude, either as a two--photon two--pion contact interaction
or in the pole terms with one vertex from
${\cal L}_2$ and the other from $\Delta {\cal L}_4$ (note that the propagator
is counted as $O(p^{-2})$).

Let us now consider the invariant amplitude for the process
$\gamma (\epsilon,k)+\pi^+(p_i)\rightarrow \gamma'(\epsilon',k')+\pi^+(p_f)$.
For a complete discussion of this process in standard chiral perturbation
theory ($\beta_1=\beta_2=0$) see Refs.\ \cite{Bijnens,Donoghue2}.
In Ref.\ \cite{Rudy} the most general renormalized irreducible off--shell
electromagnetic vertex compatible with chiral symmetry was derived
to $O(p^4)$ from Eq.\ (\ref{ls}).
The calculation of this vertex involves only Feynman diagrams which
cannot be disconnected by cutting any single internal line.
For positively charged pions and for real photons ($q^2=0, q=p'-p$)
it is found to be
\begin{equation}
\label{emv}
\Gamma^{\mu,irr}_R(p',p)=(p'+p)^{\mu}
\left(1+16\beta_1\frac{p'^2+p^2-2M^2_{\pi}}{F^2_{\pi}}\right).
\end{equation}
For $p^2=p'^2=M^2_\pi$, Eq.\ (\ref{emv}) reduces to the standard
on--shell vertex.
Furthermore, the term proportional to $\beta_1$ describes the off--shell
deviation from a structureless, pointlike vertex.
The corresponding renormalized propagator satisfying the Ward--Takahashi
identity reads \cite{Rudy}
\begin{equation}
\label{propagator}
i\Delta_R(p)=\frac{i}{p^2-M^2_{\pi}
+\frac{16 \beta_1}{F^2_{\pi}}(p^2-M^2_{\pi})^2+i\epsilon}.
\end{equation}
Note that both the electromagnetic vertex and the propagator are independent
of the parameter $\beta_2$.

For $\beta_1\neq 0$ one obtains an additional contribution of the pole
terms\footnote{Of course, using Coulomb gauge
$\epsilon^{\mu}=(0,\vec{\epsilon})$, $\epsilon'^{\mu}=(0,\vec{\epsilon'})$,
and performing the calculation in the lab frame ($p_i^{\mu}=(M_{\pi},0)$),
the additional contribution vanishes, since $p_i\cdot\epsilon=
p_i\cdot\epsilon'=0$.
However, this is a gauge--dependent statement and thus not true for a general
gauge.}
which is easily calculated with the help of Eqs.\ (\ref{emv}) and
(\ref{propagator})
\begin{equation}
\label{chp}
\Delta M_P=M_P(\beta_1\neq 0)-M_P(\beta_1=0)=
-ie^2 \frac{64 \beta_1}{F^2_{\pi}}(p_f\cdot\epsilon'\,p_i\cdot
\epsilon+p_f\cdot\epsilon\, p_i\cdot\epsilon').
\end{equation}
Clearly, one has to interpret $\Delta M_P$ of Eq.\ (\ref{chp}) as the
contribution of the off--shell form factors to the Compton scattering
amplitude in the framework of ${\cal L}'$.
This has to be compared with the standard calculation \cite{Bijnens,Donoghue2}
where the electromagnetic vertex is simply given by the pointlike vertex,
$\Gamma^{\mu,irr}_R(p',p)=(p'+p)^\mu$, i.e., there is no off--shell
contribution and the propagator is just the free propagator.

However, this is not yet the complete modification of the
{\em total} amplitude,
since the very same term in the Lagrangian which contributes to the
{\em off--shell} electromagnetic vertex also generates a two--photon contact
interaction.
In order to see this we have to insert the covariant derivative of
Eq.\ (\ref{covder}) with the external fields of Eq.\ (\ref{id}) into
Eq.\ (\ref{loff}) and select those terms which contain two powers of
the pion field as well as two powers of the electromagnetic field.
{}From the first term of Eq.\ (\ref{loff}) one obtains the following
$\gamma\gamma\pi\pi$ interaction term
\begin{eqnarray}
\label{loff1}
{\Delta\cal L}_{\gamma\gamma\pi\pi}^{(1)}&=&\frac{16\beta_1 e^2}{F^2_{\pi}}
\left(-A^2[\pi^-(\Box+M^2_{\pi})\pi^+
+\pi^+(\Box+M^2_{\pi})\pi^-]\right.\nonumber\\
&&\left.+(\partial\cdot A+2A\cdot\partial)\pi^+
(\partial\cdot A+2A\cdot\partial)\pi^-\right),
\end{eqnarray}
which translates into the contact contribution (real photons!)
\begin{equation}
\label{f1}
\Delta {\cal M}_{\gamma\gamma\pi\pi}^{(1)}=ie^2\frac{16\beta_1}{F^2_\pi}
\left(2 \epsilon\cdot\epsilon'(p^2_f+p^2_i-2M^2_\pi)
+4 (p_f\cdot\epsilon'\,p_i\cdot
\epsilon+p_f\cdot\epsilon\, p_i\cdot\epsilon')\right).
\end{equation}
The first term of Eq.\ (\ref{f1}) does not contribute to the Compton
scattering amplitude as long as the initial and final pion are on shell.
The second term precisely cancels the contribution of Eq.\ (\ref{chp}).
At first sight there appears to exist another source of a
$\gamma\gamma\pi\pi$ interaction term, namely, the second term of
Eq.\ (\ref{loff}) which gives rise to
\begin{equation}
\label{loff2}
{\Delta\cal L}_{\gamma\gamma\pi\pi}^{(2)}
=-\frac{16\beta_2 e^2 M^2_\pi}{F^2_\pi}A^2\pi^+\pi^-.
\end{equation}
{}From Eq.\ (\ref{loff2}) one obtains the Feynman rule
\begin{equation}
\label{f2}
\Delta {\cal M}_{\gamma\gamma\pi\pi}^{(2)}=-32ie^2\beta_2
\frac{M^2_\pi}{F^2_\pi} \epsilon\cdot\epsilon',
\end{equation}
and one {\em seems} to have found a modification of the Compton scattering
amplitude.
However, at this point one also has to take into account the fact that of the
two terms of Eq.\ (\ref{loff}) the second yields a modification of the
wave function renormalization constant (see Eq.\ (49) of Ref.\ \cite{Rudy})
\begin{equation}
\label{dzpi}
\Delta Z_\pi=16\beta_2 \frac{M^2_\pi}{F^2_\pi}.
\end{equation}
In order to obtain the correct result at $O(p^4)$ one has to multiply
the contact contribution of ${\cal L}_2$, $2i e^2 \epsilon\cdot\epsilon'$,
by $\Delta Z_\pi$ of Eq.\ (\ref{dzpi}).
The result precisely cancels the contribution of Eq.\ (\ref{f2}).
In fact, this cancellation has to occur since otherwise the Compton scattering
amplitude would not yield the correct low--energy Thomson limit,
$2i e^2 \epsilon\cdot\epsilon'$
(see, e.g., Ref. \cite{Bjorken}).

We emphasize that all the cancellations happen only when one consistently
calculates off--shell form factors, propagators and contact terms, and
properly takes renormalization into account.
Thus the Lagrangians of Eq.\ (\ref{ls}) which represent equivalent forms
of the same theory result in the same Compton scattering amplitude while,
at the same time, they `predict' different off--shell vertices.

\section{Conclusion}
We considered Compton scattering by a pion in the framework of chiral
perturbation theory at $O(p^4)$.
The purpose of this investigation was to clarify whether it is possible to
extract from the scattering amplitude information about the off--shell
electromagnetic vertex which enters into the calculation of the pole diagrams.
We generated an infinite class of equivalent representations of the chiral
Lagrangian by performing field transformations of the interpolating field.
This procedure allowed us to compare the results of different representations
of the same microscopic theory.

The different Lagrangians generate different off--shell electromagnetic
vertices of pions.
On the other hand, all representations result in the same Compton scattering
amplitude which we interpret to be a consequence of the equivalence theorem.
As a result of our specific example we have to conclude that even in the
framework of the {\em same} microscopic theory (in different representations)
it is not possible to uniquely extract the contributions to the scattering
amplitude which result from off--shell effects in the pole terms.
In the language of Gell--Mann and Goldberger, by a change of representation,
contributions can be shifted from class A to class B within the {\em same}
theory.
In other words, what appears to be an off--shell effect in one representation
results, for example, from a contact interaction in another representation.
It is in this sense that off--shell effects are not only
model--dependent, i.e.,
different models `predict' different off--shell form factors,
but they are also representation--dependent, that is even
different representations of the same theory `predict' different
off--shell form factors.
This has to be contrasted with on--shell S--matrix elements which,
in general, will be different for different models (model--dependent),
but always the same for different representations of the same model
(representation--independent).

In fact, we believe that our arguments are neither restricted to spin--zero
particles nor to real photons.
The discussion of off--shell effects in the literature has so far mainly been
concerned with reactions involving nucleons and virtual photons.
Even though the additional spin degrees of freedom require a more complicated
theoretical description the concepts of field transformations as discussed
above are the same.
Thus it would appear that also the off--shell electromagnetic form factors of
the nucleon as discussed in Refs.\ \cite{Naus,Tiemeijer,Song,Bos2,Doenges} are
representation--dependent and choosing a different representation of the
same microscopic model would have yielded a different off--shell behaviour.
In particular, qualitative estimates of off--shell effects, even if they are
only meant to provide an order--of--magnitude estimate, should be interpreted
with care.

We have seen that the most general result for the Compton scattering amplitude
up to $O(p^4)$ can be obtained in a representation ($\beta_1=0$) with no
off--shell effects at all in the electromagnetic vertex for real photons.
This is a special feature of the momentum expansion of chiral perturbation
theory up to $O(p^4)$, and one should not generalize this observation to
higher orders in the momentum expansion.
Higher--loop diagrams may, in general, yield off--shell contributions which
cannot be transformed away.

In conclusion, the freedom of performing field transformations allows to
shift contributions between different building blocks in different
representations of the same theory, while the on--shell S--matrix remains the
same.
In general, quantum field theoretical models will yield off--shell vertices,
however, they are not unique.
In particular, they are not only model--dependent but also
representation--dependent.

\section{Acknowledgements}

This work was supported in part by a grant from the Natural Sciences and
Engineering Research Council of Canada. One of the authors (S.\ S.) would
like to thank D.\ Harrington for stimulating discussions and for pointing
out a numerical error in an earlier version of the manuscript.
Furthermore, we would like to thank J.\ H.\ Koch for a critical reading
of the manuscript.

\end{document}